\documentclass[11pt,twoside]{article}

\def\gtsim{\mathrel{\spose{\lower.5ex \hbox{$\mathchar"218$}}
     \raise.4ex\hbox{$\mathchar"13E$}}}
\def\ltsim{\mathrel{\spose{\lower.5ex\hbox{$\mathchar"218$}}
     \raise.4ex\hbox{$\mathchar"13C$}}}
\def\aFe{[$\alpha/{\rm Fe}$]}

\def\ZH{[$Z/{\rm H}$]}

\def\apjl{ApJL}
\def\apjs{ApJS}
\def\mnras{MNRAS}

\def\aap{A\&A}

\def\araa{ARA\&A}

\def\spose#1{\hbox to 0pt{#1\hss}}

\usepackage{asp2006}
\usepackage{epsf}
\usepackage{epsfig}
\usepackage{lscape}

\begin{document}
\title{The pseudobulge of NGC~1292}   
\author{L.~Morelli$^{1}$,
 E.~Pompei$^{2}$, A.~Pizzella$^{1}$, J.~M\'endez-Abreu$^{1,3}$, E.~M.~Corsini$^{1}$, L.~Coccato$^{4}$,
 R.~P.~Saglia$^{4}$, M.~Sarzi$^{6}$ and F.~Bertola$^{1}$}   
\affil{$^1$ Dipartimento di Astronomia, Universit\`a di Padova,
  vicolo dell'Osservatorio~3, I-35122 Padova, Italy.\\
$^2$ European Southern Observatory, 3107 Alonso de Cordova,
  Santiago, Chile.\\
$^3$ INAF-Osservatorio Astronomico di Padova,
  vicolo dell'Osservatorio 5, I-35122 Padova, Italy.\\
$^4$ Max-Planck Institut f\"ur extraterrestrische Physik,
  Giessenbachstrasse, D-85748 Garching, Germany.\\
$^6$ Centre for Astrophysics Research, University of Hertfordshire,
College Lane, Hatfield, Herts AL10 9AB\\}    

\begin{abstract} 
The photometric and kinematic properties of Sb NGC~1292 suggest it
hosts a pseudobulge. The properties of the stellar population of such
a pseudobulge are consistent with a slow buildup within a scenario of
secular evolution.

\end{abstract}

\section{Introduction}

The current picture of bulge demography reveals that disk galaxies can
host bulges with different photometric and kinematic properties
\citep[see][for a review]{korken04}. Classical bulges are similar to
low-luminosity ellipticals and are thought to be formed by mergers and
rapid collapse. Pseudobulges are disk-like or bar-like components
which were slowly assembled by acquired material, efficiently
transferred to galaxy center where it formed stars.
Pseudobulges can be identified according to their morphologic,
photometric, and kinematic properties because they retain a memory of
their disky origin. Pseudobulges are expected to be more
rotation-dominated than classical bulges which are more
rotation-dominated than giant elliptical galaxies. A list of
pseudobulges characteristics was compiled by \citet{korken04}. The
more apply, the safer the classification becomes.

\section{The bulge of NGC~1292}

We derived the structural parameters, stellar kinematics, and
line-strength indices for a sample of 14 disk galaxies in
groups and clusters \citep{Moreetal08}.

\begin{figure}
\includegraphics[angle=90.0,width=0.49\textwidth,height=0.24\textheight]{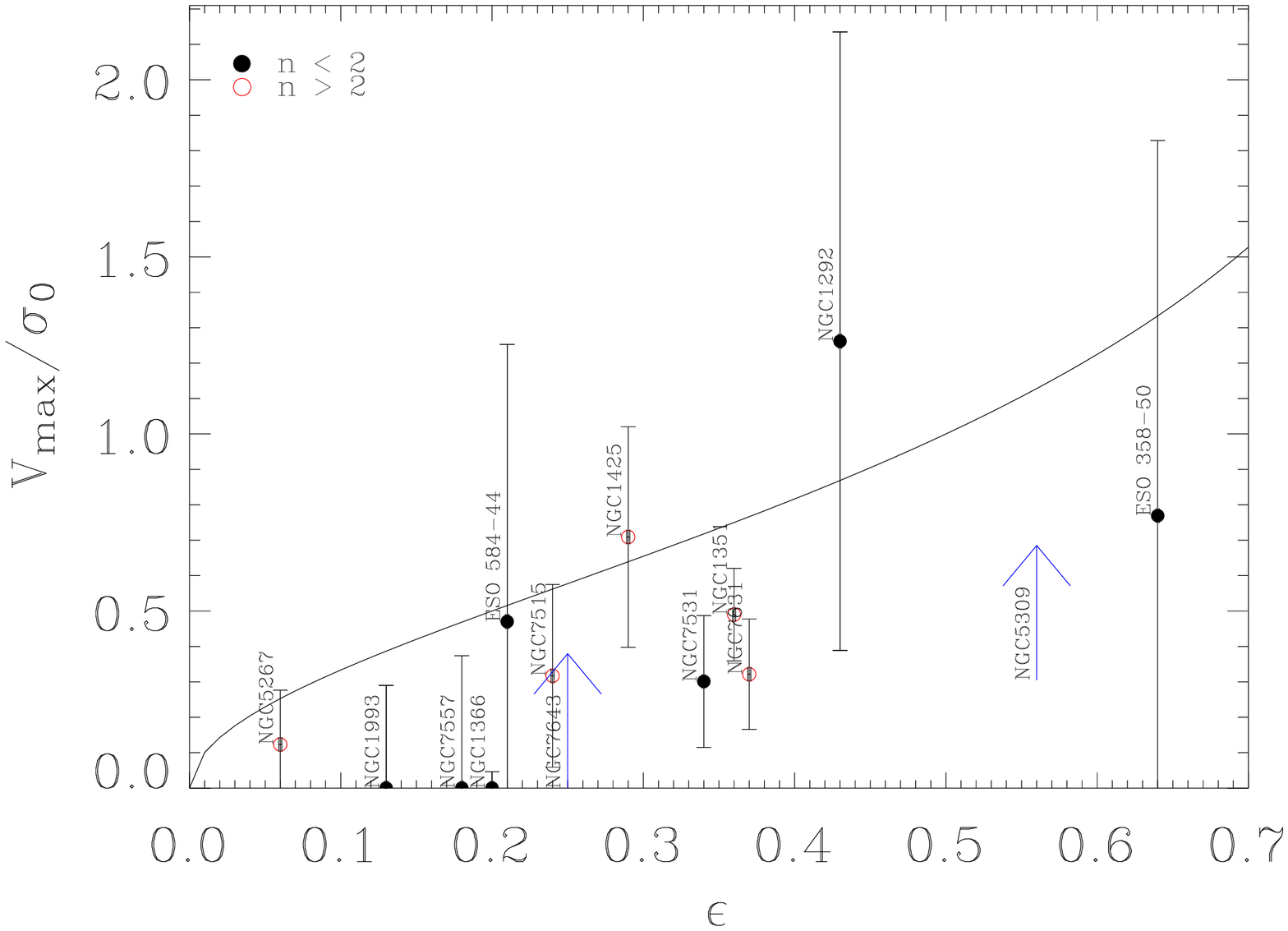}
\includegraphics[angle=90.0,width=0.49\textwidth,height=0.24\textheight]{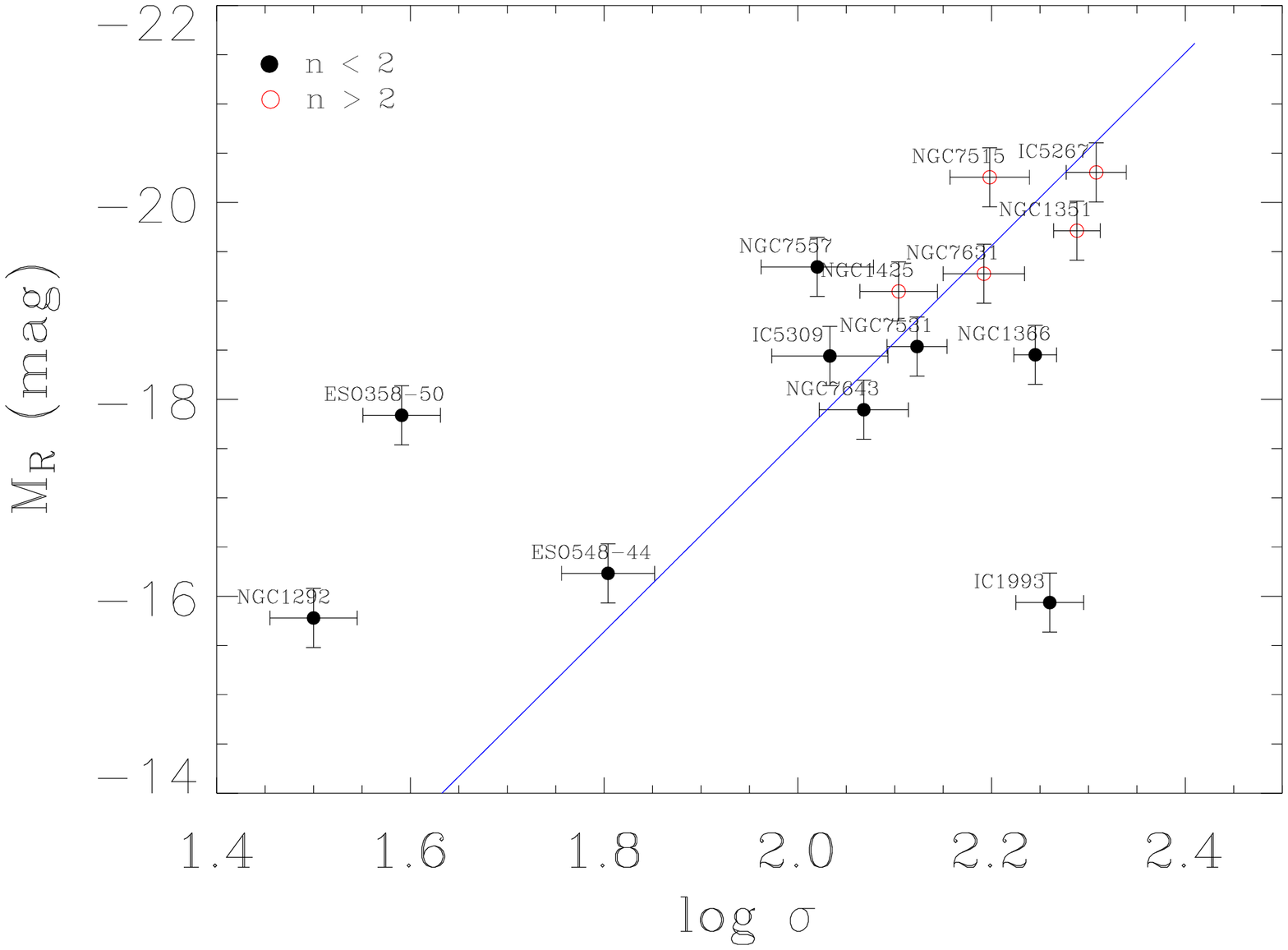}
\caption[]{Properties of the bulges studied by \citet{Moreetal08}. 
  Filled and open circles correspond to bulges with S\'ersic index
  $n\leq2$ and $n>2$, respectively.  Left panel: The location of the
  sample bulges in the $(V_{\rm max}/\sigma_0,\epsilon)$ plane.  The
  continuous line corresponds to oblate-spheroidal systems that have
  isotropic velocity dispersions and that are flattened only by
  rotation. Right panel: The location of the sample bulges with
  respect to the FJ relation by Forbes \& Ponman (1999, continuous
  line).
\label{fig:VSigma_FJ}}
\end{figure}

The bulge of the Sb spiral NGC~1292 has a S\'ersic index $n = 0.52$
and an apparent flattening which is similar to that of the galaxy disk
($q=0.6$).
The measured $V_{\rm max}/\sigma_0$ is consistent with being
significantly larger than that derived for an oblate spheroid
(Fig. \ref{fig:VSigma_FJ}, left panel).
Moreover, the bulge of NGC~1292 is not consistent with the
$R-$band Faber-Jackson relation built from
\citet[][$L\propto\sigma^{3.92}$]{forpon99} as done by
\citet{magu05}. It is characterized by a lower velocity dispersion
(or equivalently a higher luminosity) with respect to its early-type
counterparts (Fig. \ref{fig:VSigma_FJ}, right panel). NGC~1292 is also
a low-$\sigma$ (or a high-$L$) outlier with respect to the
relationship found for faint early-type galaxies by
\citet[][$L\propto\sigma^{2.01}$]{magu05}.

Information about the stellar population give more constraints on its
nature and formation process. The bulge population has a intermediate
age (3 Gyr) and low metal content (\ZH$=-0.7$ dex). The $\alpha/$Fe
enhancement is the lowest in our sample (\aFe$=-0.12$ dex) suggesting
a prolonged star formation history. The presence of emission lines in
the spectrum shows that star formation is still ongoing.

\section{Conclusions}

According to the prescriptions by \citet{korken04}, the bulge of
NGC~1292 is a good candidate to be pseudobulge. The properties of the
stellar population of such a pseudobulge are consistent with a slow
buildup within a scenario of secular evolution.


\end{document}